# Pressure induced superconductivity in MnSe

T.L. Hung[1*], C.H. Huang[1*], L.Z. Deng[2*], M.N. Ou[1], Y.Y. Chen[1] and M.K. Wu[1,3+], S.Y. Huyan[2] and C.W. Chu[2,4], P.J. Chen[5], T.K. Lee[6]

[1]Institute of Physics, Academia Sinica, Taipei. Taiwan
[2]Texas Center for Superconductivity and Physics Department, University of Houston, Houston, Texas, USA;
[3]Department of Physics, National Tsing-Hua University, Hsinchu, Taiwan
[4]Lawrence Berkeley National Laboratory, Berkeley, CA 94720
[5]Institute of Atomic and Molecular Sciences, Academia Sinica, Taipei, Taiwan
[6]Department of Physics, National Sun-Yet-Sen University, Kaoshiung, Taiwan

*TLH, CHH and LZD contribute equally to this work.
+ Email: mkwu@phys.sinica.edu.tw

**Abstract**

The rich phenomena in the FeSe and related compounds have attracted great interests as it provides fertile material to gain further insight into the mechanism of high temperature superconductivity. A natural follow-up work was to look into the possibility of superconductivity in MnSe. It was shown that MnP becomes superconducting with $T_c \sim 1$ K under pressure. We demonstrated in this work that high pressure can effectively suppress the complex magnetic characters of MnSe crystal when observed at ambient condition. MnSe under pressure is found to undergo several structural transformations: the cubic phase first partially transforms to the hexagonal phase at about 12 GPa, the crystal exhibits the coexistence of cubic, hexagonal and orthorhombic phases from 16 GPa to 30 GPa, and above 30 GPa the crystal shows a single orthorhombic phase. Superconductivity with $T_c \sim 5$ K was first observed at pressure ~12 GPa by magnetic measurements (~16 GPa by resistive measurements). The highest $T_c$ is ~ 9 K (magnetic result) at ~35 GPa. Our observations suggest the observed superconductivity may closely relate to the pressure-induced structural change. However, the interface between the metallic and insulating boundaries may also play an important role to the pressure induced superconductivity in MnSe.

**Introduction**

The rich phenomena in Fe-based superconductors [1-4] have attracted great attention because the material has offered numerous insights into the mechanism of high temperature superconductivity. The multiple-orbital nature of these materials, combined with spin and charge degrees of freedom, results in the observation of many intriguing phenomena such as structural distortion, magnetic or orbital ordering [5], and electronic nematicity [6 - 10].

The parent compounds of FeAs-based materials exhibit structural transitions from a high-temperature tetragonal phase to a low-temperature orthorhombic phase, which accompanies with an antiferromagnetic (AF) order [11, 12]. Upon doping, both the orthorhombic structure and the AF phase are suppressed and superconductivity is induced. On the other hand, FeSe undergoes a tetragonal-to-orthorhombic transition at ~ 90 K [2, 13, 14] without magnetic order at ambient pressure [14, 15] and superconductivity below ~8 K [2, 13] is crucially related to this orthorhombic distortion. The coexistence of nematic order with superconductivity without long-range magnetic order has led to arguments that the origin of the nematicity in FeSe is not magnetically but likely orbital-driven [16, 17]. More recent studies show the application of pressure leads to the suppression of structural transition, the appearance of a magnetically ordered phase at ~1 GPa [15, 18], and $T_c$ increases to a maximum of about 37 K [19–24] at ~6 GPa.

A natural follow-up work was to look into the substitution effects of Fe by other transition metals on superconductivity of FeSe. We reported that substitution of up to 6% Mn to Fe does not affect much the superconductivity in FeSe [25]. On the other hand, only 3% Cu-substitution to Fe completely suppressed the superconductivity of FeSe. It is known that MnSe forms in a cubic structure at ambient condition and exhibits anomalous magnetic structure [26, 27] so that no superconductivity could be detected. The NiAs-type FeSe favors to form hexagonal $\gamma$-Fe$_{1-x}$Se that exhibits both antiferromagnetism and ferrimagnetism depending on composition [28]. And superconductivity only exists when FeSe forms tetragonal structure [2]. Therefore, it will be valuable to investigate whether one could manage to form MnSe with crystal symmetry favorable for superconductivity.

MnP, which has an orthorhombic structure (with Pbmn symmetry), was found to be the first Mn-based superconductor with transition temperature ~1 K under 8 GPa [29]. It is noted that at ambient pressure MnSe exhibits very much the same magnetic behaviors [30] as those observed in MnP [29]. Therefore, it is of great interest to investigate whether superconductivity can also be induced in MnSe system. An idea to test such a possibility is to use the smaller ion sulfur to replace selenium to generate internal pressure. Thus, we have carried out the detailed structural study of Mn(Se-S) system [30].

Based on the refined lattice parameters of the Mn(Se-S) system [30], we estimated the equivalent compression pressure (E.C.P) in MnSe by systematic sulfur substitution, using the third-order Birch-Murnaghan equation of state reported by Catherine McCammon [31]. The results suggest the E.C.P. of MnS (relative to MnSe) is ~13.2 GPa as shown in **Supplementary Table 1**. The estimated E.C.P. of MnS relative to MnSe is higher than that required to induce superconductivity in MnP compound. However, no superconductivity was observed in MnS as it maintains in cubic phase with an antiferromagnetic-like order at ~150 K and an anomalous ferromagnetic-like order at ~ 16 K. Nevertheless, the results demonstrated that partial substitution of Se by S could indeed effectively suppress the partial transformation of the cubic phase to hexagonal phase [30].

Wang et al. showed in their study of MnSe that the lattice collapsed under high pressure [32]. In their studies, the crystal structure of MnSe distorts to orthorhombic phase with space group Pnma under ~ 30 GPa [32]. And this orthorhombic phase is identical to the MnP superconducting phase. They also showed the compound is in low spin states under pressure based on X-ray emission spectroscopy and the transport measurement on MnSe indicated the sample becomes metallic at ~ 30 GPa. However, the temperature and pressure range in their study were rather limited. Therefore, it is desirable to carry out a more detailed investigation on MnSe over a wide temperature range under high pressure.

**Experimental Methods**

Polycrystalline MnSe samples were prepared by solid-state reaction method using raw materials of Mn (99.95%, Alfa-Aesar), and Se (99.95%, Acros-Organic). Angle dispersion X-ray diffraction (ADXRD) experiments were performed using a symmetric diamond anvil cell (DAC) with 300 μm culets. A rhenium gasket was pre-indented to a thickness of ~70 μm from an initial thickness of 250 μm. A 150 μm-diameter sample chamber was drilled in the center of the pre-indented gasket. Micro-meter ruby balls were placed inside the sample chamber as the pressure gauge [45]. Helium gas was used as a pressure transmitting medium using a gas loading system. ADXRD measurements were collected using the beamline BL01C2 at the National Synchrotron Radiation Research Center (NSRRC), Taiwan. The X-ray energy was 19 keV.

High pressure resistivity measurements used a DAC with 400 μm culets. A rhenium gasket was covered by cubic-BN powders for insulating the electrical leads. A MnSe of dimension ~70 μm × 70 μm × 15 μm is loaded in a sample chamber filled with hexagonal-BN as pressure transmitting medium. Gold foils were used as electrodes to connect the sample and gold wires. Details of the contacts arrangement are shown in Fig. S1. Pressure was determined by ruby florescent method [45]. Resistance measurements at low temperature was measured by Van der Pauw method in the $^4$He cryostat.

A mini-DAC fabricated from BeCu alloy, which was adapted into a Quantum Design Magnetic Property Measurement System (MPMS), was used for ultrasensitive magnetization measurements under high pressures [33]. A pair of 300-μm-diameter culet-sized diamond anvils was used. The gaskets were made from nonmagnetic Ni–Cr–Al alloy. Each gasket was pre-indented to ~20 μm in thickness, and a ~120-μm-diameter hole was drilled to serve as the sample chamber. The mixture of methanol and ethanol in a ratio of 4:1 was used as the pressure transmitting medium. The applied pressure was measured by the fluorescence line of ruby powders. A piston-cylinder-type high-pressure cell, compatible with MPMS, was used when performing low pressure measurements up to 1.3 GPa, where the pressure medium was Daphne-7373 oil and the pressure manometer was a lead piece.

**Computational Method**

The first-principles calculations are performed using Quantum Espresso [34] with norm-conserving local-density approximation pseudopotentials. The energy cut-off for the plane-wave expansion is 60 Ry. At low pressures MnSe is semiconducting, so Coulomb $U$ = 5 eV is included to deal with the correlation of the localized 3$d$ orbitals. At higher pressures, both hexagonal and orthorhombic MnSe are metallic. The Coulomb $U$ is not included because it is unimportant in metallic systems where the orbitals are more delocalized.

**Results and Discussions**

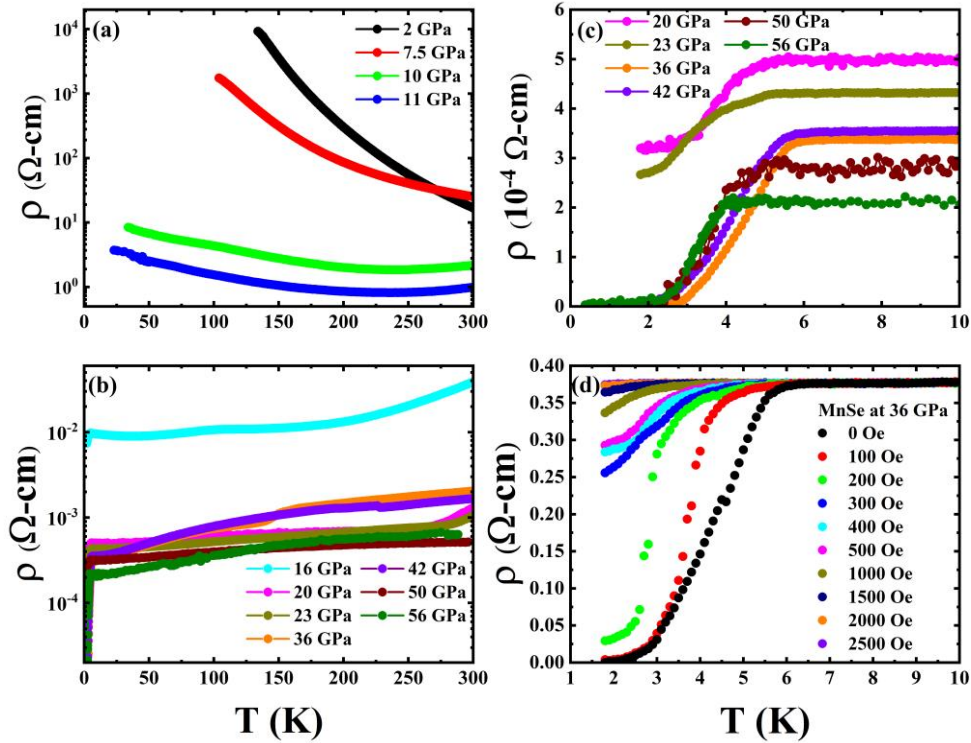

Figure 1. Temperature dependence of resistivity of MnSe at different pressures (a) below 11 GPa; and (b) above 16 GPa where the resistance drop at low temperature appears. (c) Shows the detailed resistive transition below 10 K, and (d) displays the resistive transition under different magnetic fields confirms the superconducting transition.

Fig. 1(a) and (b) shows the temperature dependence of resistivity for MnSe at different pressures. The result at ambient pressure is consistent with the previous report [30]. In the pressure range lower than 16 GPa, the sample exhibits semiconducting behavior. However, the semiconducting gap value decreases with increasing pressure. An abrupt drop in resistivity at room temperature was observed at ~ 10 GPa, and a second resistivity drop appears at pressure ~16 GPa, at which the sample changes to metallic behavior. Meanwhile, a small drop in resistance is observed at ~ 4 K above 16 GPa. This low temperature drop becomes more prominent as pressure increases, showing clear superconducting transitions above 20 GPa. It is noted that a third drop in resistivity at room temperature occurs above 20 GPa and the data above 30 GPa show much larger residual resistance ratio. Fig. 1(c) presents the detailed resistive transition of the sample with pressure above 20 GPa, showing the superconducting transition with zero resistance above 2 K at pressures above 36 GPa. Fig. 1(d) displays the resistive transition under different magnetic fields. The magnetic field effect further confirms the superconducting transition nature.

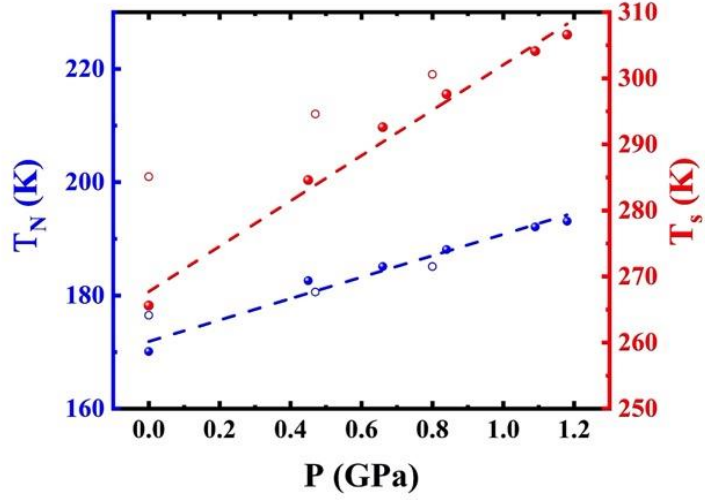

Figure 2. $T_N$ vs. P and $T_s$ vs. P under pressures up to 1.2 GPa. Solid symbols represent results during loading process while open symbols show data collected during unloading process.

In magnetic measurements, we first reproduced the observations reported by Huang *et al*. [30] on MnSe at ambient condition in the pressure cell. Two magnetic anomalies were clearly detected in MnSe. One broad anomaly occurs between 100 K and 200 K with a signal at the order of $10^{-4}$ emu/Oe · g, which was suggested to be a coupling results of the magnetic locking effect of β-MnSe and thermal fluctuation on the short-range ferromagnetic sheets in α-MnSe. The other anomaly around 266 K was attributed to a partial transformation of the cubic-phase to hexagonal-phase. Here we defined the peak position for the anomaly between 100 K and 200 K as $T_N$ and the peak position of the anomaly above 250 K as $T_s$. Both $T_N$ and $T_s$ increased as pressure increased up to 1.2 GPa, with $dT_N/dP \sim 18.9$ K/GPa and $dT_s/dP \sim 34.3$ K/GPa, as shown in Fig. 2, which is different from the doping effect by replacing Se with S [30]. (Details of the measured results can be found in Fig. S3). As pressure increases, the amplitude of first anomaly firstly increases and then decreases, and then increases again, while the second anomaly firstly increases and then decreases.

During unloading, our results showed that $T_N$ was more reversible than the $T_s$ (Fig. S4). After fully releasing the pressure, by comparing the result with zero pressure data before applying pressure, we found out $\Delta T_N \sim 15$ K while $\Delta T_s \sim 30$ K. We would like to point out during unloading a third anomaly appeared between 240 K and 280 K (Fig. S4). The peak position of this anomaly decreases as pressure decreases.

For the high-pressure magnetization measurements using DAC, we first measured the sample with zero pressure, and two anomalies were both detectable (green line in Fig. 3a), though the signal size is small due to the small mass of the sample. At 2.68 GPa, the two anomalies were almost completely suppressed, while we observed an up-turn at lower temperature, which is similar to the results for $MnSe_{1-x}S_x$ at lower temperature [30], but no anomaly at around 150 K was observed at pressure below 11 GPa. As we continued to increase the pressure, a diamagnetic drop was observed at 11.75 GPa and above, as seen in Fig. 3b which indicates possible pressure-induced superconductivity. The amplitude of the diamagnetic drop also increased as pressure increased. It is noted that a small hump was detected at pressure between 11.75 GPa and 25.92 GPa at ~ 150 K, Fig. 3(c), which is similar to the antiferromagnetic (AFM) transition reported in the $MnSe_{1-x}S_x$ [30]. Above 25.92 GPa, the data suggests that the pressure suppresses the AFM transition, meanwhile the diamagnetic transition becomes more prominent, as shown in Fig.4.

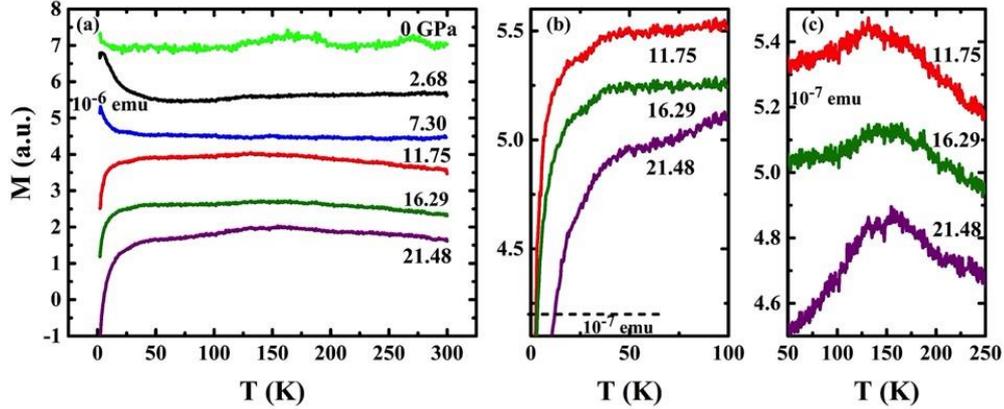

Figure 3. M. vs. T at different pressures up to 21.48 GPa.

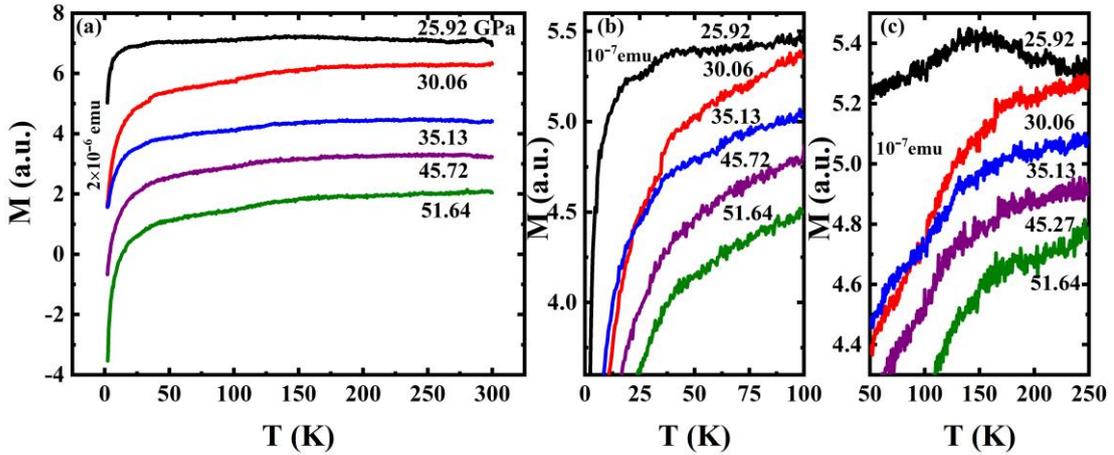

Figure 4. M. vs. T at different pressures between 25.92 GPa and 51.64 GPa.

Two kind of pressure dependence of MT and RT measurements were used to determine the superconducting transition temperature ($T_c$) (shown in Figs. S5 & S6 and Fig. S2 respectively). $T_c$ vs. P obtained from M(T,P) data is summarized in Fig. 5a. A local minimum point appears around 26 GPa, which is consistent with a phase transition to be discussed later. We also include in the same figure $T_c$ vs. P determined by resistive measurements. The results exhibit certain inconsistency in the $T_c$ value, with the magnetic measurements showing relatively higher $T_c$s and richer behavior. It is noted that the superconducting transitions observed at various pressures are generally broad. However, the onset $T_c$, from either RT or MT measurements, as explained earlier, are well-defined. Thus, the observation of higher $T_c$ onset by magnetization measurements is an experimental fact. The difference of the $T_c$ values derived from resistive and magnetic susceptibility measurements most likely was due to the pressure inhomogeneity with different DAC cells (from two different labs) and pressure medium used. Additionally, observation of zero resistance depends on having a percolative path of superconductivity across the sample; whereas, a diamagnetic response only needs a shell with thickness of order a penetration depth around isolated structural domains or grains.

The effect of pressure homogeneity on superconducting is a complex issue. For example, Matsubayashi *et al.* [35] reported that superconductivity in $Bi_2Te_3$ was very sensitive to hydrostatic condition of the applied pressure and demonstrated that the superconducting phase could only survive under strong uniaxial stress. Another example is the pressure-induced superconductivity in $CaFe_2As_2$, which could only be observed above 0.5 GPa by measurements using organic medium [36]; on the contrary, no superconductivity was observed in a helium medium [37]. Miyoshi et al. reported in FeSe system, the increase in $T_c$ is suppressed under non-hydrostatic pressure [38]. And much earlier work by Klotz and Schilling found that $T_c$ of Bi2212 is suppressed faster under hydrostatic conditions [39].

We used hexagonal-BN powder as the pressure medium for R(T) measurements. Inevitably we expect to generate uniaxial stress inside the cell. This characteristic was reflected in the broadening of the ruby R1 peak [40], as shown in Fig. S8. The M(T) measurements used the mixture of methanol and ethanol in a ratio of 4:1 as pressure medium, which is expected to exhibit better hydrostatic condition. The results of detailed analysis of the ruby-R1 broadening in RT measurements are shown in Figure S8.

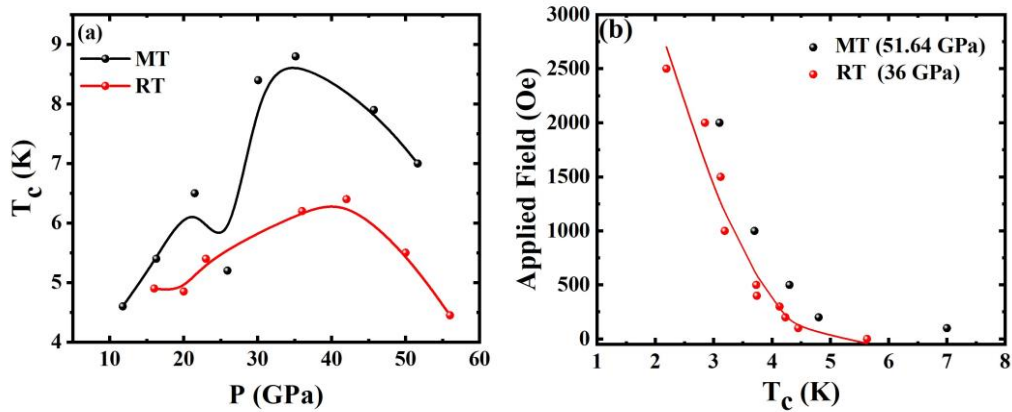

Figure 5. (a) $T_c$ vs. P from both MT and RT results. For MT measurement, $T_c$ is determined from MT as described in supplementary information. For R(T) measurements, $T_c$ is taken at the onset of resistive transition. The $T_c$ is obtained by the intersection of two adjacent lines fit of $d\rho/dT$ (b) Field dependence of $T_c$ obtained from MT (at 51.64 GPa) and RT (at 36 GPa).

To further confirm the superconducting transition, we measured the sample resistivity at 36 GPa and the magnetization at ~52 GPa under different magnetic fields. The detailed measured results are shown in Figures 1(c) and S7. It clearly demonstrates that magnetic field suppresses the transition, confirming its superconducting nature. Fig. 5b displays the magnetic field dependence of $T_c$ for both RT at 36 GPa and MT at ~52 GPa. The results further confirm the superconducting nature of the magnetic and resistive transitions. The extrapolated upper critical field observed is consistently ~8100 Oe. Using the Ginzburg-Landau formula, $\mu_0 H_{c2}(0) = \Phi_0/2\pi\xi^2$, the estimated superconducting coherence length is $\xi \sim 201$ Å.

To gain more insight into the origin for the observed pressure-induced superconductivity, we performed a series of X-ray diffraction measurements on MnSe under pressures. The results are presented in Fig. 6. The crystal structure of MnSe is rock salt (cubic, $Fm\bar{3}m$) with lattice constant 5.4697 Å at ambient condition. The results clearly show that MnSe undergoes two structural transformations at 12.2 GPa and 30.5 GPa, respectively. The diffraction pattern exhibits the coexistence of cubic phase and hexagonal phase at 12.2 GPa, which is slightly higher than that reported by McCammon who showed the partial transformation at 9 GPa [31]. A surprising observation, which was not reported by earlier study, was the sample to exhibit another partial transition at ~ 16 GPa. The new phase present is orthorhombic so that within the pressure ranges from 16 GPa to 30 GPa the sample is in a mixed state with the coexistence of cubic, hexagonal and orthorhombic phases, as shown in Fig. 7. The X-ray patterns show another structural transformation at 30 GPa that MnSe completely transformed to orthorhombic phase (MnP-type, Pnma) with lattice $a = 5.7527$ Å, $b = 3.1045$ Å, and $c = 6.0434$ Å. These data are consistent with the results from theoretical calculations though our refinement gave slightly larger lattice parameters.

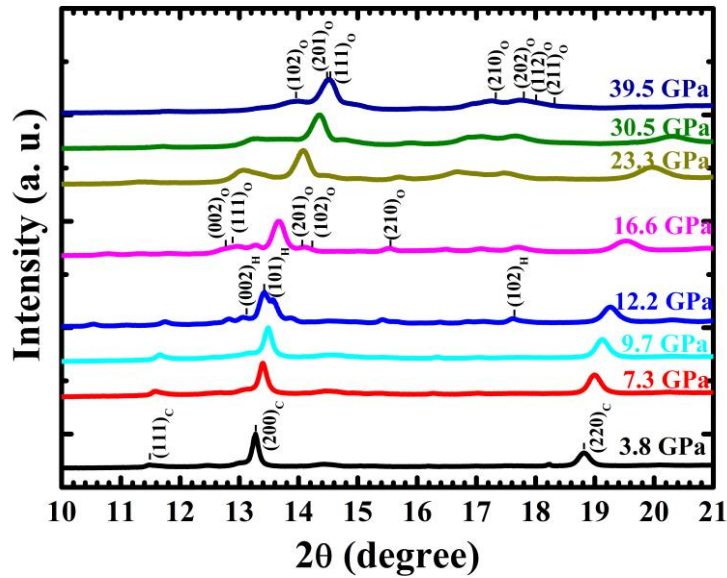

Figure 6. In situ synchrotron XRD patterns of MnSe during compression at room temperature.

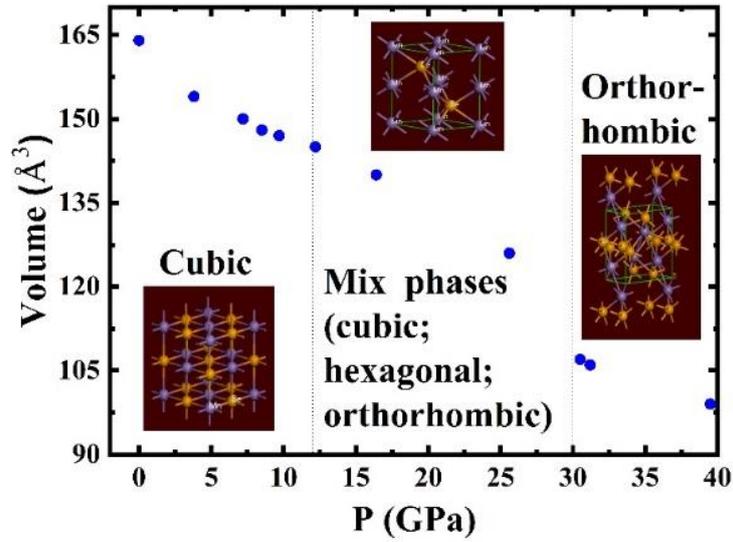

Figure 7. The pressure dependence of volume for MnSe. The initial structure is cubic phase, and the intermediate structure retains cubic phase and exhibits the new hexagonal and orthorhombic phases, and the final structure is orthorhombic phase, which could be a new superconducting phase.

At ambient pressure, our calculations indicate MnSe is cubic with AFM configuration consistent with that reported [30]. Besides, a structural phase transition from hexagonal to orthorhombic MnSe at pressure ~ 40 GPa is found, which agrees with the experimental observation. The crystal and band structures of the orthorhombic phase at 40 GPa are shown in Fig. 8. It is also noted that MnSe shows the low-spin state ($S = 0.5$) in the orthorhombic phase as reported [27].

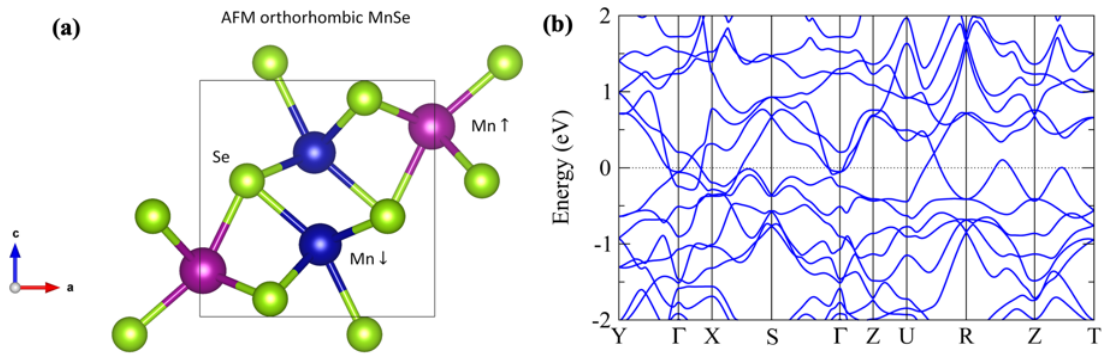

Fig. 8. (a) The crystal structure of orthorhombic MnSe. Purple, blue, and green spheres represent Mn↑, Mn↓, and Se atoms, respectively. (b) The band structure of orthorhombic MnSe with $a = 5.280$ Å, $b = 2.999$ Å, and $c = 5.472$ Å.

The partial cubic-to-hexagonal transformation at 12 GPa could be similar to the low temperature stress induced transformation of 30% cubic to hexagonal structure at ambient condition, which has been extensively investigated and is understood to be the source for anomalous magnetic observed in MnSe at ambient condition [30]. The hexagonal phase remains to be insulating so that it could not be the source for the observed pressure-induced superconductivity. The observation of the orthorhombic phase appears at ~16 GPa coincides with the appearance of metallic behavior from the resistive measurements. It is also the pressure at which the onset of superconducting resistive transition is identified, though the magnetization measurements suggest the transition appears at lower pressure (~12 GPa). The magnetic measurements suggest the existence of two superconducting domes, one from 12 GPa to 25 GPa and the other dome with maximum $T_c$ appears at ~ 35 GPa. Unfortunately, the resistive measurements do not show the same results.

Based on the results of high-pressure X-ray diffraction studies one would suggest that pressure-induced superconductivity is connected to the observed orthorhombic phase appeared at high pressure. If this were the case, one would expect the diamagnetic signal after 30 GPa would substantially increases as the material become orthorhombic single phase. However, the observed diamagnetic signals below and above the structural transition are comparable. Furthermore, one would also expect the $T_c$ values determined by MT and RT results after 30 GPa would be the same if superconductivity is associated with the orthorhombic phase. In fact, the observed results show that the difference in onset $T_c$ by two different methods is even larger above 30 GPa.

Normally, one would expect to observe the resistive onset $T_c$ to be either the same or higher than that obtained from magnetization measurements. The observation of a higher $T_c$ by magnetic measurements in pressurized MnSe is rather unusual. It has been reported in a K-doped FeSe superconductor ($K_2Fe_{4+x}Se_5$ system) that the magnetic transition temperature is higher than the resistive transition temperature [41]. For example, both the magnetization and resistive measurements show consistently an onset $T_c$ ~ 31 K for the samples with x = 0.2 prepared with rapid quenching directly after annealing at 850 C displaying high superconducting volume fraction. However, for the sample after post-annealed at 400 C for 2-hour, the $T_c$ determined by magnetization measurement remains with an onset at 31 K but with much smaller volume fraction, the resistive transition was suppressed to a $T_c$ onset at 21 K. The longer the low temperature annealing time results in the smaller superconducting volume fraction and the lower resistive superconducting transition $T_c$. This result was due to the presence of a high volume of non-conducting phase in the post-annealed sample.

It is noted that diamagnetic susceptibility up to 45 K can only be observed in the ac susceptibility at high frequency in ultrathin FeSe films due to the possible interface-enhanced superconductivity [43]. The observation of relatively low upper critical field in MnSe might provide additional support to the picture of interfacial effect as the observed transition could be due to the Josephson junction coupling between grains [44].

In summary, we have undoubtedly demonstrated the pressure-induced superconductivity in MnSe. The anomalous magnetic behavior of MnSe at ambient condition was quickly suppressed by applying pressure. Superconductivity kicks in at ~12 GPa as shown by magnetic measurement (and ~ 16 GPa by resistive measurement). The appearance of the resistive transition $T_c$ coincides nicely with the appearance of orthorhombic phase. The transition temperature in MnSe under pressure is much higher than that of the pressure-induced superconductivity in MnP [9] though they may exhibit the same crystalline phase under high pressures. A local minimum point appears around 26 GPa by magnetic measurements. Though it is very possible that the pressure-induced superconductivity is associated with the pressure-induced orthorhombic phase, however, our data suggest that the interfacial effect between the metallic and insulating boundaries may play an important role to the induced superconductivity.

## Acknowledgments

The authors at Academia Sinica thank the financial support from the Academia Sinica Thematic Research Grant No. AS-TP-106-M01 and the Ministry of Science and Technology of Taiwan Grant No. MOST108-2633-M-001-001. The work performed at the Texas Center for Superconductivity at the University of Houston is supported by the U.S. Air Force Office of Scientific Research Grant FA9550-15-1-0236, the T. L. L. Temple Foundation, the John J. and Rebecca Moores Endowment, and the State of Texas through the Texas Center for Superconductivity at the University of Houston.## References